# Synthesis of Silver Colloids: Experiment and Computational Model

Ionel Halaciuga, Daniel T. Robb,* Vladimir Privman, and Dan V. Goia

*Center for Advanced Materials Processing, Clarkson University, Potsdam, NY 13699, USA*
**Department of Physics, Astronomy, and Geology, Berry College, Mount Berry, GA 30149, USA*

## Abstract

We summarize our recent results [1] that model the formation of uniform spherical silver colloids prepared by mixing iso-ascorbic acid and silver-amine complex solutions in the absence of dispersants. We found that the experimental results [2] can be modeled effectively by the two-stage formation mechanism used previously to model the preparation of colloidal gold spheres [3]. The equilibrium concentration of silver atoms and the surface tension of silver precursor nanocrystals are both treated as free parameters, and the experimental reaction time scale is fit by a narrow region of this two-parameter space. The kinetic parameter required to match the final particle size is found to be very close to that used previously in modeling the formation of uniform gold particles, suggesting that similar kinetics governs the aggregation process. The model also reproduces semi-quantitatively the effects of temperature and solvent viscosity on particle synthesis.

## 1. Introduction

Highly dispersed uniform particles are widely used presently in various areas of technology and medicine. While theoretical and computational modeling of chemical synthesis of colloids and nanoparticles has advanced recently, more progress needs to be made in understanding the complex phenomena involved, which take place over multiple length and time scales. Here we report computational modeling of the size distribution of secondary polycrystalline colloid particles produced by aggregation of primary nanosize crystalline precursors. Although initially colloidal dispersions of narrow size distribution were widely thought to consist of monocrystalline particles [4], experimental evidence [5-8] showed that, in many cases, the final particles are in fact polycrystalline. Since aggregation of diffusing particles generally produces a widening size distribution, explanation of the narrow size distribution presented a theoretical challenge. Several aggregation models that could produce narrow size distributions were proposed [3,9,10].

In particular, a model which yielded size distributions with narrow (relative) width by a two-stage process, coupling initial nucleation of primary nanocrystals with the aggregation of these nanocrystals into secondary particles, was developed [3]. In this model, the initial fast nucleation of nanocrystals leads to a sizeable peak of small secondary aggregates. If the nucleation rate of nanocrystals then decreases in time according to a proper protocol, as occurs naturally in the nucleation process, then the initial peak of the secondary aggregate distribution grows to larger sizes, while few additional secondary aggregates are formed.

Thus a well-defined peak of secondary particles grows with its average size increasing much more rapidly than its width, resulting in a narrow size distribution. An important simplification was to approximately account for the growth of primary particles after nucleation, as well as the absorption of monomers (atoms, molecules) by the secondary particle distribution, with the assumption that the aggregating primary particles were of an (experimentally determined) uniform size throughout the process. In addition, from experimental observations, the secondary particles were assumed to compactify rapidly into spheres with the bulk density, on time scales much faster than that of the diffusive processes. This allowed the Smoluchowski rate [11] $K_s = 4\pi R_s D_1$ for the diffusive capture of smaller particles with diffusion constant $D_1$ by a larger particle with radius $R_s$, to be used as a basis for the aggregation rates of primary particles. The Smoluchowski rate corresponds to an assumption of a diffusion-limited process, i.e., "instantaneous" reaction/merging of the smaller particles with the larger particle.

In general, with $N_s(t)$ denoting the concentration of secondary particles containing $s$ primary particles, the rate equations for aggregation of secondary particles can be written

$$\frac{dN_s}{dt} = \sum_{j=1}^{\lfloor s/2 \rfloor} f_{s-j,j} K_{s-j,j} N_{s-j} N_j - \sum_{i=1}^{\infty} f_{i,s} K_{i,s} N_i N_s, \quad (1)$$

where $\lfloor s/2 \rfloor$ refers to the greatest integer less than or equal to $s/2$, and where the first sum is understood to be zero for $s=1$. The Smoluchowski rate for two spherical particles is defined [11] more precisely as $K_{ij} = 4\pi (R_i + R_j)(D_i + D_j)$, with, under the instantaneous reaction assumption, the kinetic coefficients $f_{ij} = 1$ for $i \neq j$, but $f_{ij} = 1/2$ for $i = j$. However, it has been observed experimentally and argued theoretically that larger secondary particles almost never merge [3], presumably because of the relatively small area of potential contact between their surfaces (and other kinetic factors), suggesting that $f_{ij}$ can be taken to be zero for $i, j$ both above a certain size. In addition, the merging of small primary particles (generally ~ 10 nm) does not fully comply with the assumptions behind the Smoluchowski rates, both because the concept of a depletion zone becomes suspect in the case of small equally-sized particles, and because one can easily imagine that encounters between such particles may not always result in a merger (violating the instantaneous reaction assumption). Thus, the smallest secondary particles may also require adjusted kinetic factors $f_{ij}$.

Here we consider experimental data for synthesis of spherical silver colloid particles. The experimental details were reported in [2], and pertinent results are reproduced here. We use the model of only singlet (precursor nanocrystal) capture by larger aggregates [11], which corresponds to taking $f_{ij} = 0$ unless $i = 1$ or $j = 1$. This assumption of singlet dominated aggregation, which will be explained shortly, was also made in the first application of the two stage model to the formation of gold spheres [3]. For simplicity, we will refer to the parameter $f_{1,1}$ as $f$.

Our aim in this work is to further develop and test the model, as well as establish that the newly studied silver colloid system is described by the same approach previously used [3]. Initially [3], the model produced a distribution with small relative width, but with an average size 3-5 times too small, indicating that too many secondary particles were produced. A second modeling effort for the gold system [12] demonstrated that reducing the kinetic parameter $f$ to values significantly less than 1, as relaxing the instantaneous reaction assumption for singlet-singlet aggregation would suggest, reduced the number of produced secondary particles and increased the average size of the final secondary particle distribution.

The paper is organized as follows. In Sec. 2, we summarize the experimental procedure and results recently reported [2] for the production of uniform silver spheres. In Sec. 3, we describe the two-stage model for particle growth previously applied to understand the synthesis of gold spheres, using a highly accelerated integration scheme [1] to approximate the evolution of the average size of the particle distribution in time. In Sec. 4, we apply the model to the production of silver colloids, performing a simultaneous fit of the surface tension and equilibrium concentration parameters. We then compare the value of the kinetic parameter required to match the experimental size distribution to the value required in the gold system, and study the effect of solution viscosity and temperature.

## 2. Experimental System

We apply the two-stage model here to chemical synthesis of uniform silver spheres via reduction of silver-polyamine complexes with iso-ascorbic acid. This experimental system, described in detail recently [2], has commercial relevance as it is a convenient way to produce uniform dispersed silver particles that are useful for the production of silver conductive structures used in the electronics industry. Whereas many methods for the preparation of silver particles are accessible, the precipitation in homogeneous solutions remains the most flexible approach due to the possibility of using an ample range of solvents, reductants, dispersants, and complexing agents.

Depending on the experimental conditions chosen, the experimental system investigated in [2] yields spherical silver particles of a narrow size distribution, with an average diameter ranging from 80 nm to 1.3 μm. The prime benefit of this approach is that, in contrast to most precipitation methods, it generates dispersed silver spheres in concentrated systems in the absence of dispersants. As a result, the particles are free of the organic residues that adversely affect their performance in most electronic applications. From the modeling point of view, the dispersing agent should not impact the dynamics of the aggregation of nanosize precursors, because the precursors (as well as the monomers) are below the size range affected by such agents. The results presented in this paper, specifically the similar value of the kinetic parameter, $f$, required to achieve agreement between the computed and experimental average particle size in the gold and silver systems, support this expectation.

As in the previously studied gold system [3], here both electron microscopy and X-ray diffraction have shown that the silver spheres were composed of smaller crystalline subunits. The field emission scanning electron microscopy (FESEM) images corresponding to the $T = 60$ C case in [2], revealed the presence of subunits with an approximate size of 12-20 nm. The size of the subunits was also calculated from X-ray diffraction measurements using the Scherrer formula [13]. The calculated size of 18-20 nm agreed with the size observed by FESEM, providing strong evidence that the final silver spheres formed by an aggregation process and that the two stage model should be applicable.

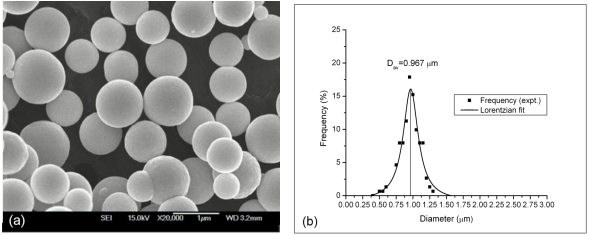

**Figure 1.** (a) FESEM image of spherical silver particles produced at 60C by the method of [2]. (b) Histogram of particle diameters taken from the FESEM images.

The succession of vivid colors during the early stages of the precipitation confirmed the presence of the dispersed subunits and growing aggregates. An electron micrograph of the silver particles produced at $T = 60$ C is shown in Fig. 1, as well as their size distribution obtained by a direct measurement of over 200 particles from FESEM images. In [2], the effects of varying a number of experimental parameters on the reduction process were examined. In Sec. 4, we will evaluate the extent to which the two-stage model can explain the results in solvents of different viscosity and at varying temperature.

## 3. Two-Stage Model of Particle Growth

The two-stage model for particle growth was developed to explain the narrow size distribution occurring in synthesis of gold spheres [3,12]. The model was motivated by experimental observations [3] that the uniform gold particles were actually polycrystalline, formed by aggregation of many smaller crystalline subunits. In the model, an initial process of nucleation of clusters (primary particles) is described by standard burst-nucleation theory. The reduction process results in a supersaturated solution with monomer (here, aqueous neutral metal atom) concentration $c$. Driven by thermal fluctuations, small nanoclusters (embryos) are produced, whose size distribution is controlled by the free energy of an $n$-monomer embryo, of the following form [3,12],

$$\Delta G(n,c) = -nkT \ln(c/c_0) + 4\pi a^2 n^{2/3} \sigma, \qquad (2)$$

where $k$ is the Boltzmann constant, $T$ is the temperature in Kelvin, $c_0$ is the equilibrium concentration of monomers, and $\sigma$ is the effective surface tension. The first term in Eq. (2) is the free energy contribution of the "bulk" of the embryo. The second term represents the surface free energy, proportional to the surface area of the embryo, and therefore to $n^{2/3}$. The effective solute radius $a$, chosen so that the radius of an $n$-solute embryo is $an^{1/3}$, is defined by requiring that $4\pi a^3/3$ is the "unit cell" volume per monomer (including the surrounding void volume) in the bulk material crystal structure.

At small cluster sizes, the surface term dominates, creating a free-energy barrier to nucleation of a thermodynamically stable cluster. The peak of the nucleation barrier occurs at the critical cluster size,

$$n_c = \left[\frac{8\pi a^2 \sigma}{3kT \ln(c/c_0)}\right]^3. \qquad (3)$$

For $n < n_c$, the concentration $P(n,t)$ of $n$-monomer nanoclusters is assumed to follow a thermal distribution,

$$P(n,t) = c \exp\left[\frac{-\Delta G(n,c)}{kT}\right]. \qquad (4)$$

The rate of production of supercritical clusters is then given [3] by

$$\rho(t) = K_{n_c} c P(n_c,t) = K_{n_c} c^2 e^{-\Delta G(n_c,c)/kT}, \qquad (5)$$

where $K_n$ is the rate constant in the Smoluchowski description of diffusive capture of particles [11]. These expressions involve various approximations described in earlier works [3,12].

A key quantity is the rate at which monomers are consumed by the growing supercritical clusters,

$$\frac{dc}{dt} = -n_c \rho(t). \qquad (6)$$

thereby reducing the supply of monomers available to constitute the thermal distribution, Eq. (4). The nucleation rate $\rho(t)$ from Eq. (5) is then used as input to a kinetic model for the aggregation of the primary particles into secondary, polycrystalline aggregates. The master equation for the aggregation process under the assumption of singlet-dominance takes the form

$$\frac{dN_s}{dt} = K_{s-1} N_1 N_{s-1} - K_s N_1 N_s, \quad \text{for } s > 2, \quad (7)$$

where $N_s(t)$ is the time-dependent number density (per unit volume) of secondary particles consisting of $s$ primary particles (the expressions for $s = 1, 2$ will be given shortly).

The attachment rate constant, $K_s$, is modeled using the Smoluchowski rate for diffusive capture [11], of the form $K_s = 4\pi (R_1 + R_s)(D_1 + D_s)$. Here $R_s = 1.2 r s^{1/3}$ is the radius of a secondary particle containing $s$ primary particles, with the factor 1.2 calculated as $(0.58)^{-1/3} \simeq 1.2$, where 0.58 is the typical filling factor for random loose packing of spheres [14]. The diffusion constant of a secondary particle containing $s$ primary particles is given by $D_s = D_1 s^{-1/3}$. The average radius $r$ of the primary particles is available experimentally (from X-ray diffraction), from which the primary particle diffusion constant can be calculated using the Stokes-Einstein relation [3]. Note that in [3], the attachment rate was approximated as $K_s = 4\pi R_s D_1$, corresponding to the limit $s \gg 1$. Here we follow Ref. 12 and use the full expression for $K_s$. For the case $s = 2$, the master equation was taken in [12] to be

$$\frac{dN_2}{dt} = f K_1 N_1^2 - K_2 N_1 N_2. \quad (8)$$

As discussed in the Introduction, for the assumption of instantaneous reaction (diffusion limited kinetics), the kinetic parameter $f$ would be 1/2. Here, using the singlet-dominated approach of Eq. (7), we will eventually take $f \ll 1/2$ to limit the number of secondary particles produced. The evolution of the singlet ($s = 1$) population is given by

$$\frac{dN_1}{dt} = \rho(t) - \sum_{j=2}^{\infty} j \frac{dN_j}{dt}. \quad (9)$$

To make the model numerically tractable, the diffusive growth of already-nucleated primary particles is not explicitly included, but is incorporated by the use of the experimentally determined primary particle size. In combination with Eq. (6), this can lead to the conservation of matter being violated. Thus, the final distribution $N_s(t)$ is regarded as relative [3,12], and must be normalized to correspond to the total amount of matter initially present, given by the initial monomer concentration, $c(t = 0)$.

## 4. Simulation Results for Silver

In applying the two-stage model to the experimental Ag system described in Sec. 2, we found that the available literature values [15] for equilibrium concentrations of Ag (in highly purified, neutral, degassed water) were in the range $c_0 = 1.4\text{-}2.0 \times 10^{20}$ m$^{-3}$. Given the large difference from the value for gold [3], and the fact that surface oxides and adsorbed O$_2$ may affect the solubility [14], we decided to treat not only the surface tension but also the equilibrium concentration as a free parameter. A series of simulations were run at different values of the two parameters $\sigma, c_0$, using the kinetic parameter value $f = 0.5$ [1].

The results are illustrated in Fig. 2, which combines a contour plot of the saturation time with a representation of the average particle sizes (at full saturation) on a regular grid of points $\sigma, c_0$.

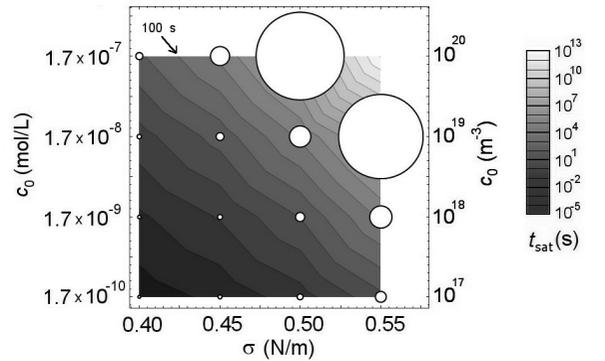

**Figure 2.** Contour plot of the saturation time.

Interestingly, the saturation time depends in a systematic way on both the surface tension and the equilibrium concentration. It varies over an enormous range, from $10^{-5}$ s in the lower left corner of the figure to $10^{13}$ s in the upper right corner. We caution, however, that the assumption of the dominance of the singlet-cluster attachment process becomes suspect for saturation times above ~ 300-500 s, at which point the processes of cluster-cluster attachment and Ostwald ripening become relevant. Thus the saturation time of approximately $10^{13}$ s for parameter values $\sigma = 0.55 \text{ N/m}$, $c_0 = 10^{20} \text{ m}^{-3}$, while accurate for the mathematical model, would likely be much lower in experiment. For the particles shown in Fig. 1, the experimentally observed time scale for saturation was ~ 100 s. This time scale can be reproduced in simulations by using values of $\sigma$, $c_0$ drawn from anywhere along the $t = 100$ s contour in Fig. 2.

A previous study [16] of the surface tension of silver nanoparticles reported the surface tension in colloidal solution as $\sigma = 0.525 \text{ N/m}$. We therefore provisionally identify the joint parameter values $\sigma = 0.525 \text{ N/m}$, $c_0 = 1.15 \times 10^{18} \text{ m}^{-3}$, which result in a saturation time $t_{\text{sat}} = 96.7 \text{ s}$, as the best fit to our experimental results. Our main finding at this point however is not that these parameter values are precisely correct, but that the joint values of $\sigma$, $c_0$ must be drawn from the relatively narrow $t = 100 \text{ s}$ contour region. The particle size distribution as a function of time, for best-fit parameter values and kinetic parameter $f = 0.5$, is shown in Fig. 3 (left data series).

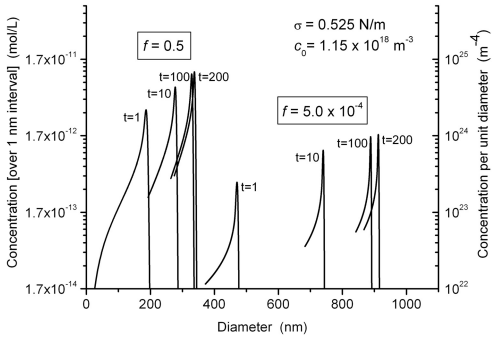

**Figure 3.** Time evolution of particle size distribution for the best fit parameters.

The average size at full saturation is smaller than the experimentally observed particle size, but using kinetic parameter $f = 5.0 \times 10^{-4}$ leaves the saturation time practically unchanged, while it increases the average size at full saturation to quantitative agreement with the experimental results. This value of the kinetic parameter, required to match the experimentally observed final particle size in this silver system, is quite close to the value $f = 1.0 \times 10^{-4}$ which was found, in [3], to match the experimental size in the gold system. In addition, this value is equal to that fitted, in [7], to match the experimental particle size in CdS systems. This suggests that the kinetic parameter represents similar physics at work in the three systems.

We next investigate the degree to which the two-stage model reproduces the effects of varying the solvent used in synthesizing the Ag particles. As shown in [2], the use of a more viscous solvent than water, namely diethylene glycol (DEG), resulted in a much smaller final particle size. The saturation time was observed to be quite close to the aqueous system (~ 100 s). We assume that the diffusion constants of both individual silver atoms and primary particles in DEG are related to those in water, $D_{\text{DEG}} = \left( \eta_{\text{H}_2\text{O}} / \eta_{\text{DEG}} \right) D_{\text{H}_2\text{O}}$. To achieve a saturation time ~ 100 s, it is then necessary to decrease slightly the parameter values, to $\sigma = 0.505 \text{ N/m}$ and $c_0 = 1.0 \times 10^{18} \text{ m}^{-3}$, which is consistent with the fact that alcohols are usually found to have lower equilibrium metal-atom concentrations than water.

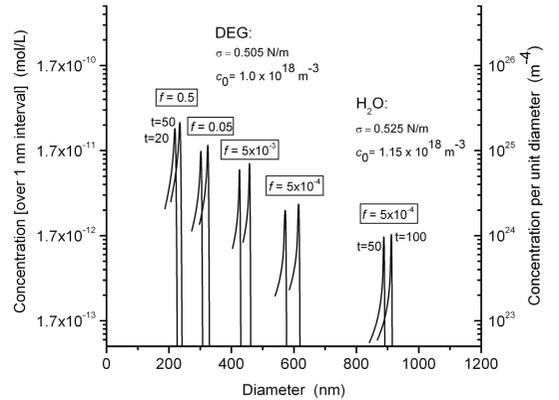

**Figure 4.** Time evolution of particle size distribution in different solvents.

Furthermore, it is reasonable to suppose that the kinetic coefficient $f$ in DEG is somewhat larger than that in water, since the increased viscosity of DEG may result in a higher probability of attachment on encounters between singlets (primary particles): Indeed, rotational diffusion was identified as a contributing factor to the suppression of the formation of dimers in protein crystallization in solutions [17]. As shown in Fig. 4, by increasing the kinetic parameter from $f = 5 \times 10^{-4}$ to $f = 0.05$, the average size at saturation is brought closer to the experimental value of 80 nm. Still, the remaining factor of 3-4 indicates that our model does not completely capture the behavior in the DEG solvent. Resolving this discrepancy may require a more accurate understanding of the kinetic coefficients $f_{ij}$ in Eq. (1).

Finally, we investigated the effect of the temperature on the final size distribution, attempting to model the experimental results reported in [2], using the best-fit parameters, and the value of the kinetic parameter $f = 5 \times 10^{-4}$. The calculated average diameters are in reasonable agreement with those from experiment. The saturation times show the same trend as in the experiment, decreasing as the temperature is increased from 40 C to 80 C, but they vary over a larger range in simulation than in experiment. The actual simulated size distributions at each temperature are shown in Fig. 5.

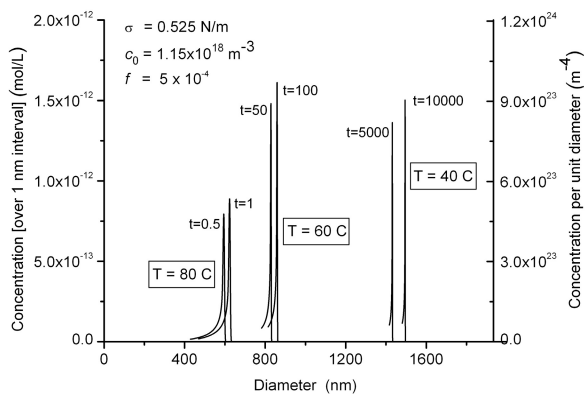

**Figure 5.** Time evolution of particle size distribution at different temperatures.

## 5. Conclusion

We have used the two-stage model previously applied to gold and cadmium sulfide systems to model the production of uniform spherical silver particles. The use of an accelerated integration scheme [1,18] enabled the exploration of the parameter space of both $\sigma$ and $c_0$, with the conclusion that they play a comparable role in determining the saturation time for particle production. We determined that the kinetic parameter $f$ in the singlet-dominated aggregation scheme does not noticeably affect the saturation time in the model, and that the value of $f$ required to match the average final particle size in the Ag system is very similar to that required in the Au and CdS systems. This provides further evidence that the model captures important aspects of the kinetics of aggregation in chemical synthesis across different experimental systems. With parameters optimized for 60 C in aqueous solution, the model was found to account semi-quantitatively for experimental results at different temperatures and in the solvent diethylene glycol. It seems likely that a more accurate model for the kinetics of aggregation at small primary particle size, intermediate between the singlet-dominated approach with $f \ll 1/2$ and an approach [19] involving a cutoff size for primary particle aggregation, may be required to produce quantitative agreement with experiment over a range of temperatures and solvents.

This research was supported by the NSF under grant DMR-0509104.

## References

[1] D. T. Robb, I. Halaciuga, V. Privman and D. V. Goia, *submitted for publication* (2008).
[2] I. Halaciuga and D. V. Goia, *J. Mater. Res.* **23**, 1776 (2008).
[3] V. Privman, D. V. Goia, J. Park and E. Matijević, *J. Col. Int. Sci.* **213**, 36 (1999).
[4] V. K. LaMer, and R. J. Dinegar, *J. Am. Chem. Soc.* **72**, 4847 (1950).
[5] D. Murphy-Wilhelmy and E. Matijević, *J. Chem. Soc. Faraday Tr.* **80**, 563 (1984).
[6] E. Matijević and P. Scheiner, *J. Col. Int. Sci.* **63**, 509 (1978).
[7] M. Ocana, C. J. Serna and E. Matijević, *Col. Polym. Sci.* **273**, 681 (1995).
[8] M. P. Morales, T. Gonzales-Carreno and C. J. Serna, *J. Mater. Res.* **7**, 2538 (1992).
[9] W. T. Scott, *J. Atm. Sci.* **25**, 54 (1968).
[10] J. Th. G. Overbeek, *Adv. Col. Int. Sci.* **15**, 251 (1982).
[11] R. v. Smoluchowski, *Z. Phys. Chem.* **29**, 129 (1917).
[12] J. Park, V. Privman and E. Matijević, *J. Phys. Chem. B* **105**, 11630 (2001).
[13] A. L. Patterson, *Phys. Rev.* **56**, 978 (1939).
[14] R. German, *Particle Packing Characteristics* (Metal Powder Industries Federation, Princeton, 1989).
[15] *American Institute of Physics Handbook*, 3rd ed., edited by D. E. Gray (McGraw-Hill, New York, 1972).
[16] S. F. Chernov, Y. V. Fedorov and V. N. Zakharov, *J. Phys. Chem. Solids* **54**, 963 (1993).
[17] C. N. Nanev, *Crystal Growth Design* **7**, 1533 (2007).
[18] V. Privman, *Mater. Res. Soc. Symp. Proc.* **703**, Article T3.3, 577 (2002).
[19] S. Libert, V. Gorshkov, V. Privman, D. Goia and E. Matijević, *Adv. Col. Int. Sci.* **100-102**, 169 (2003).